\documentstyle[12pt]{article}
\begin{document}
%\font\bsf=cmbx12 scaled\magstep 0
\font\bss=cmr12 scaled\magstep 0
\title{The G\"odelizing Quantum-Mechanical Automata}
\author{ A.V. Yurov
\small\\ Theoretical Physics Department, \small\\ Kaliningrad
State University, Russia,  yurov@freemail.ru}
\date {}
\renewcommand{\abstractname}{\small Abstract}
\maketitle
%\begin{document}
\maketitle
\begin{abstract}
Using Albert results we argue that  we don't need new physics  to
understand G\"odelization. Albert quantum automaton can
"understand" both a formal system and a G\"odel proposition which
can't be obtained within this system. There are two significant
conclusions. The first speaks "against Penrose" whereas the
second speaks for him.
\end{abstract}
\thispagestyle{empty}
\medskip

\section {Introduction}

In his two books [1] Penrose has argued that we need new physics
in order to understand the "mind". The crucial point of his
approach is the G\"odel theorem and the natural ability of our
mind to do the G\"odel procedure which will be referred as {\em
G\"odilization}. What G\"odel showed [2] was how to transcend any
system of formalized rules.  And because our mind can do it, than
it cannot be the "formal system" itself. But our physics is
nothing but some (very complex and intricate) "formal system" so
it should be incomplete and we need new physics which must
include our mind in.

This is Penrose point, but I don't think so (with the proviso in
last Section). In this report I argue that quantum-mechanical
automaton do something which can be called as G\"odelization. The
argument below is far from being a rigorous proof but I believe
that it makes it rather plausible. The crucial point will be
Albert approach to quantum-mechanical automata [3]. What Albert
showed was that "there are some combination of facts that can in
principle be predicted by an ({\bf quantum-mechanical} - A.Y.)
automaton {\em only about itself}" [3]. I'll show that we can
interpret it as an quantum automaton's ability to G\"odelize, so
we don't need new physics and "new mind" to understand the
mystery of G\"odelization.

\section{G\"odel theorem and quantum automaton}

Let $p_n(w)$ is the propositional function applied to number $w$,
where $n$ will be called further the G\"odel number. The strings  of propositions
which constitute a proof of some theorem (for example, $p_n(w)$,
if this statement is true in our system)  be $U(x)$, where $x$ is
the G\"odel number of the proof. We denote our (classical) formal
system as $W_c$.

The fundamental G\"odel proposition has the form
$$
p_w(w)=\sim\,\exists\, x\,\left[U(x)\,\,{\rm
proves}\,\,p_w(w)\right], \eqno(1)
$$
where $\sim$, $\exists$ are usual logical "quantifiers": $\exists$
mean "there exists...such that", $\sim$ mean "it is not so
that...".

What G\"odel showed was that (1) can be strongly written in the
special formal system $W_c$. It's clear that $p_w(w)$ must be
true as an arithmetical proposition (we suppose that $W_c$ is
good, reconcilable system and there are not proofs of false
statements) and therefore, by the  implication of the statement
(1), it can't be proves within the system $W_c$.

It is seems that G\"odel proposition (1) can be constructed in
any formal system therefore any formal system will be limited by
the G\"odel theorem but it is not the case for us because we can
G\"odelize and, consequently, transcend any system of formalized
rules.

Now let consider the quantum machine which must proves theorems
in formal system $W_c$. Let $S$ is quantum system with Hilbert
space $H_S$. We introduce observable $P_k(w)$ such that for any
$|w;k\rangle_S\in H_S$ it will be true that
$$
P_k(w) |w;k\rangle_S=k|w;k\rangle_S. \eqno(2)
$$
So eigenvalues of $P_k(w)$ will be G\"odel numbers of the
propositions $p_k(w)$. In the same Hilbert space we define
logical  "gaits" $U_j$. The entire  list of all such "gaits" will
be referred as quantum formal system $W_q$. We'll say that
quantum formal system $W_q$ can proves the proposition $p_k(w)$
(about number $w$) if such chain of "gaits" exist
$$
U= U(w,k)\equiv \prod_j\,U_j,
$$
that
$$
|\psi\rangle_S\to |w;k\rangle_S=U|\psi\rangle_S, \eqno(3)
$$
for some initial state $|\psi\rangle_S\in H_S$. In this case we
can start out from $|\psi\rangle_S$ to find $|w;k\rangle_S $;
then to measure the average value of $P_k(w)$, to obtain the
G\"odel number $k$
$$
k={}_S\langle\psi|U^+P_k(w)U|\psi\rangle_S,
$$
which allow one to reconstruct the proposition $p_k(w)$. It may
not be simple task but we can do it as a matter of principle.

When this procedure is realizable one? To understand this we
suppose that the quantum formal system $W_q$ is universal  one
and it is suitable for testing $|w;k\rangle_S$ to be solution of
the (2). Substituting $|w;k\rangle_S$ in place of $
|\psi\rangle_S$ into the (3) we must obtain the same vector
~\footnote{May be one get another vector $|w;m\rangle_S$ with
$m\ne k$. In this case we can redefine $U$ to exclude this
outcome.}
$$
U|w;k\rangle_S=u(w;k) |w;k\rangle_S, \eqno(4)
$$
so it must be
$$
\left[P_k(w),U\right]=0. \eqno(5)
$$
The equation (5) is at one with uncertainty relation. Indeed, let
the quantum formal system $W_q$ is instructed to measure (or
calculate: it is the same in our case)	the value of $p_k(w)$.
The observable of $W_q$ is $U$ so both $P_k(w)$ and $U$ must be
measurable simultaneously. This is another way to obtain (5).

What about G\"odel proposition (1)? It is propositional function
applied to number $w$ with G\"odel number $w$ which can't be
obtained (proved) in formal systems $W_c$ and $W_q$. Write
$P_w\equiv P_w(w)$. It would be
$$
\left[P_w,U\right]\ne 0. \eqno(6)
$$
Otherwise we can use $U$ to find $|w\rangle_S\equiv |w;w\rangle_S$
and, after all, to calculate G\"odel proposition $p_w(w)$. But it
is impossible because of the G\"odel theorem.
\section{Albert quantum automaton and auto-description}
Let consider the automaton $A$ which is instructed to measure and
to record the value of $P_w$ and $U$. It being known that
$$
U|\psi\rangle_S=u|\psi\rangle_S,\qquad |\psi\rangle_S=\sum_w
c_w|w\rangle_S,\qquad P_w|w\rangle_S=w|w\rangle_S, \eqno(6)
$$
where $P_w$, are observables which are comparing to G\"odel
propositions $p_w$.  Let us take the state of the system $S$ to
be  $|w\rangle_S$.  When measurement of the $P_w$ is finished,
the state of the composite system $S+A$ will be [3]
$$
|w^{(1)}\rangle =|w\rangle_S \bigotimes|w\rangle_{P_w}, \eqno(7)
$$
where $ |w\rangle_{P_w}\in H_A$, $H_A$ is the Hilbert space of
the automaton $A$ and  $|w\rangle_{P_w}$ is the eigenvector of
the "Albert observable" $G(P_w)$ (in [3] Albert has used another
notation),
$$
G(P_w) |w\rangle_{P_w}={\tilde w} |w\rangle_{P_w}. \eqno(8)
$$
This observable measures the value predicted by the automaton for
the $P_w$. The prediction will {\em accurate} if
$$
E(P_w) |w^{(1)}\rangle=\left({\tilde w}-w\right)
|w^{(1)}\rangle=0, \eqno(9)
$$
with
$$
E(P_w)\equiv G(P_w)-P_w. \eqno(10)
$$
Below we'll restricting the accurate predictions (${\tilde w}=w$).

Albert observables do commute with the rest
observables~\footnote{We stress that $G(P_w)$ is not the function
of the observable $P_w$ as well as $G(U)$ is not function of
$U$.},
$$
\left[G(P_w),P_w\right]= \left[G(P_w),U\right]=
\left[G(P_w),G(U)\right]= \left[G(U),P_w\right]=
\left[G(U),U\right]=0,
$$
so
$$
\left[E(P_w),E(U)\right]=\left[P_w,U\right]\ne 0, \eqno(11)
$$
what is at one with the uncertainty relations and, at the same
time, with the G\"odel theorem. Indeed, a G\"odel proposition
$p_w$ has no proof within formal system $W_q$  so the predictions
about $U$ and $P_w$ can't both be accurate. Thus we have the same
situation as above. If that's the case, why did we introduce the
second quantum automaton $A$? This because Albert has
demonstrated that quantum automaton $A$ can transcend the
restriction (11) in a certain sense. To do it the automaton must
measure something about {\em itself}.

Let us take the initial state of the system $S$ to be the
eigenvector of $U$:  $|\psi\rangle_S$. It mean that $W_q$ is "in
action" and $P_w$ is beyond the system. In spite of this the $A$
is instructed to measure $P_w$. When this measurement is finished
the state of composite system $S+A$ will be
$$
|\psi^{(1)}\rangle=\sum_w c_w|w^{(1)}\rangle, \eqno(12)
$$
because of linearity of quantum-mechanical equations. In (12)
$|w^{(1)}\rangle $ are defined by the (7). Now one can introduce
new observable $U^{(1)}$ such that
$$
U^{(1)}|\psi^{(1)}\rangle= u^{(1)}|\psi^{(1)}\rangle. \eqno(13)
$$
What is the $U^{(1)}$? We can call it  the "gait" of new formal
system $W_q^{(1)}$ which is, partially, recorded in the memory of
the automaton $A$. It's clear that
$$
\left[P_w,U^{(1)}\right]\ne 0, \eqno(14)
$$
so its seems that the new formal system $W_q^{(1)}$ is not more
powerful  system  that initial	one ($W_q$). But now,  if the
automaton will measure the observable $U^{(1)}$ then we get
$$
|\psi^{(1)}\rangle\to |\psi^{(2)}\rangle =
|\psi^{(1)}\rangle_{U^{(1)}}\bigotimes |\psi^{(1)}\rangle,
\eqno(15)
$$
(see (7)), where
$$
G(U^{(1)}) |\psi^{(1)}\rangle_{U^{(1)}}=u^{(1)}
|\psi^{(1)}\rangle_{U^{(1)}}, \eqno(16)
$$
therefore
$$
E(P_w) |\psi^{(2)}\rangle=E(U^{(1)}) |\psi^{(2)}\rangle=0,
\eqno(17)
$$
thought $P_w$ and $U^{(1)}$ do not commute (see (11))!

Thus the automaton in the state $|\psi^{(2)}\rangle$ can predict
accurate both $P_w$ and $U^{(1)}$. I believe we can call it
"Godelization" in a sense. It is not mean that the automaton can
proves $p_w$ using within $W_q^{(1)}$. What is implied by this
that $P_w$  (and therefore the proposition $p_w$) and $U^{(1)}$
are jointly satisfiable for the automaton $A$. In other words,
the automaton $A$ can "understand" both a G\"odel proposition
$p_w$ and a formal system $W_q^{(1)}$. The classical automaton
can't do it. To force classical automaton "understand" $p_w$  and
$W_c^{(1)}$ we need to load it with more powerful "formal system"
${\tilde W}_c$. It is not the case when we deal with quantum
automaton.

An last but not least. The G\"olelization above is the personal
file of the automaton $A$. The observable $U^{(1)}$ and observable
$P_w$ are external ones for the another automaton ${\tilde  A}$,
so it can't to 'understand'  both $U^{(1)}$ and $P_w$. To
G\"odelize it must find its {\em personal state} $|{\tilde
\psi}^{(2)}\rangle$.
\section{Conclusion}
There are two significant conclusions. The first speaks "against
Penrose" whereas the second speaks for him.
\newline
\newline
{\em Against}. We don't need new physics  to understand
G\"odelization. Albert quantum automaton can "understand" both a
formal system and a G\"odel proposition which can't be obtained
within this system.
\newline
\newline
{\em For}. If G\"odelization is unalgorithmic procedure then, at
least, we can admit that quantum mechanics really containing
"something unalgorithmic".
\newline
\newline
\centerline{\bf References} \noindent
\begin{enumerate}
\item R. Penrose {\em New Emperor's Mind}, Oxford University
Press, (1989); R. Penrose {\em Shadows of the Mind}, Oxford
University Press, (1994).
\item K. G\"odel, {\em \"Uber formal unentscheidbare S\"atze per Principia
Mathematica und verwandter System I}. Monatshefte f\"ur Mathematik
und Physic, 38, 173-98 (1931).
\item D. Albert, {\em On Quantum-Mechanical Automata.}.\rm\, Phys.
Lett. A {\bf 98}, 249-52 (1983).

\end{enumerate}

\end{document}